# Mobile Safety Application for Pedestrians

**Sukru Yaren Gelbal, Mustafa Ridvan Cantas, Bilin Aksun Guvenc, and Levent Guvenc**

**Gopichandra Surnilla and Hao Zhang**

## Abstract

Vulnerable Road User (VRU) safety has been an important issue throughout the years as corresponding fatality numbers in traffic have been increasing each year. With the developments in connected vehicle technology, there are new and easier ways of implementing Vehicle to Everything (V2X) communication which can be utilized to provide safety and early warning benefits for VRUs. Mobile phones are one important point of interest with their sensors being increased in quantity and quality and improved in terms of accuracy. Bluetooth and extended Bluetooth technology in mobile phones have enhanced support to carry larger chunks of information to longer distances. The work we discuss in this paper is related to a mobile application that utilizes the mobile phone sensors and Bluetooth communication to implement Personal Safety Message (PSM) broadcast using the SAE J2735 standard to create a Pedestrian to Vehicle (P2V) based safety warning structure. This implementation allows the drivers to receive a warning on their mobile phones and be more careful about the pedestrian intending to cross the street. As a result, the driver has much more time to safely slow down and stop at the intersection. Most importantly, thanks to the wireless nature of Bluetooth connection and long-range mode in Bluetooth 5.0, most dangerous cases such as reduced visibility or No-Line-of-Sight (NLOS) conditions can be remedied.

## Introduction

VRU safety is an important problem concerning all other road users. The existing high number of traffic accidents with a large number of VRU injuries and fatalities [1] indicate that this topic is still a significant concern today. In order to address this concern and to make the roads safer for VRUs, better active and passive safety measures are being extensively researched all over the world by automotive companies and academic research groups. V2X connectivity is one of the most widely used means of providing safety and better driving experience to drivers and pedestrians. Dedicated Short-Range Communications (DSRC) technology [1] is one of the most popular means of communication along with Cellular V2X (C-V2X) for information transfer between vehicles under Vehicle to Vehicle (V2V) and between vehicles and infrastructure under Vehicle to Infrastructure (V2I) connectivity. Within the V2X research, pedestrian safety is also considered and researched under Vehicle to Pedestrian (V2P) connectivity [2], [3], [4]. Publications in this field have a wide variety of technologies for information transfer [5, 6, 7] and a wide variety of pedestrian and driver models and analysis methods [8, 9, 10, 11] to calculate the possible future safety violation accurately. There is also research work that focuses on collision avoidance and collision free path planning once pedestrian location is determined to be on a colliding path with the vehicle [12], [13].

Since availability of Bluetooth 5.0 technology in recent mobile phone models getting wider, this technology is also becoming a viable means of communication between VRUs and other road users. Mobile phones can be carried by any VRU on hand or in pocket, while having wide array of sensors available for measurement, tracking or predicting the movement and behavior of the VRU, typically a pedestrian. With Android API, development of an Android app utilizing this communication is possible, where different modes of communication under Bluetooth can provide different benefits, extended advertisement (CODED PHY) being the most advantageous in terms of larger message size and communication distance. Using the sensors available on the phone along with the Bluetooth advertisement capabilities, information can be transferred between road users, enabling the implementation of safety applications that can conveniently be installed in mobile phones that may reside in a car or be carried by a VRU.

This paper discusses the implementation and testing of a mobile safety application for pedestrians where the Bluetooth extended advertisement feature is utilized to transfer Pedestrian to Vehicle P2V information using the SAE J2735 Personal Safety Message (PSM) structure. Both the drivers and pedestrians can use this application in their mobile phones and benefit from it. Implementation was designed as different modules which are discussed in the following sections. This discussion and results presented are aimed towards demonstrating the capability and potential of this implementation and use case in terms of VRU safety.

The remainder of the paper is organized as follows. The P2V Over Bluetooth section discusses the communication between vehicle and pedestrian as well as the Android Application Programming Interface (API) and message structure used along with what type of information is conveyed. This is followed by the Simple Collision Prediction section where the paper discusses the base scenario configuration created for the implementation, definition of the variables and calculation of the possible future collision. Afterwards, warning design, severities and calculations are discussed in the Driver Warning section. Test scenarios are discussed and results are demonstrated in the Testing and Results section. The paper ends with the Conclusions and Future Work section where a summary discussion of the main results are presented along with recommendations for improvements that can be realized in the future.

## P2V Over Bluetooth

In order to calculate the possible collision and show the driver a warning, pedestrian location and motion information needs to be transferred to vehicle. By doing this communication using wireless technology, we also ensure that even if there is No-Line-of-Sight (NLOS) and sensors cannot detect the pedestrian, the vehicle would know that there is a pedestrian(s) and be aware of their movement information such as location, speed and heading. Therefore, one of the main enablers of this implementation is the capabilities and relatively wide availability of Bluetooth 5.0 in recent Android phone models. Moreover, extended advertisement capability in recent mobile Bluetooth implementations allows users to transfer larger chunks of data over longer distances. Although extensive analysis on advantages and disadvantages of using extended advertisement over normal operation is not within the scope of this paper, extended advertisement was preferred for the implementation and testing due to increased communication distance in both Line-of-Sight (LOS) and NLOS situations.

For the message structure, PSM from the SAE J2735 standard [14] was preferred to carry the pedestrian localization and motion information. Along with having location, speed and heading fields for calculating the collision possibility and warning, PSM also has fields such as user type, device use state, cross request and cluster size, attachment. All of these fields are useful for future improvements of the algorithm regarding behavior prediction, pedestrian group clustering and more accurate collision prediction in cases where the VRU has attachments such as pets, carts or wheelchair.

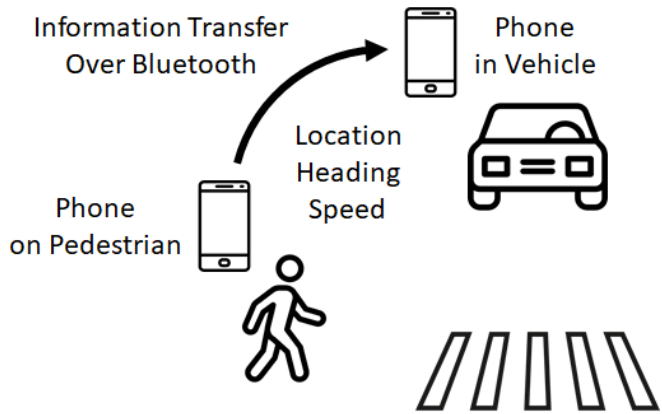

Figure 1. Illustration of P2V over Bluetooth.

In our current implementation, the developed algorithm requires location, heading and speed information from the pedestrian, which are obtainable using the GPS sensor on the phone. Therefore, a software program was prepared using the Bluetooth Low Energy (BLE) module in Android API 28 [15]. This software has the capability to encode, transmit, receive and decode PSM messages through extended advertisement, depending on the user selected. Since only the driver is warned in this implementation, the pedestrian is the transmitting side and the vehicle is the receiving side. This P2V communication is illustrated in Figure 1. Populated fields for PSM are the location (latitude and longitude), heading and speed, as shown in the figure. At the end, the phone in the vehicle obtains information about both the pedestrian location and motion, through communication, and the vehicle, through the GPS sensor of the phone in the vehicle.

## Simple Collision Prediction

Collision prediction calculations are handled on the phone in the vehicle, after receiving the PSM messages from the pedestrian phone. The base scenario geometry defined assumes the vehicle and pedestrian are moving straight, similar to one of the configurations in the National Highway Traffic Safety Administration (NHTSA) high safety risk crash scenarios as illustrated in Figure 2.

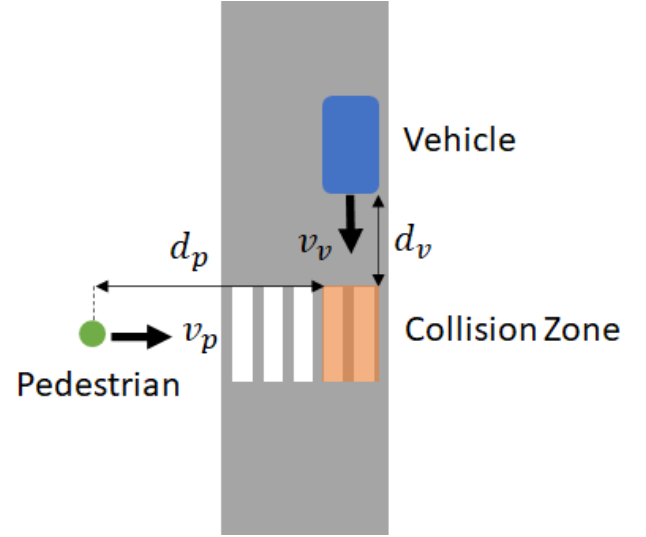

Figure 2. Geometrical configuration and variables for base scenario.

A collision zone is defined at the crosswalk as a rectangular area where the pedestrian and vehicle headings intersect. Vehicle speed is denoted by $v_v$, pedestrian speed is denoted by $v_p$ and distances to collision zone are denoted respectively as $d_v$ and $d_p$. This configuration is created as a base scenario and will be the only one used in the scope of this paper. However, after other parts of the design and implementation are established, the scenario configuration can be modified, if desired, to calculate the prediction for different cases such as right and left turns. Using the variables mentioned above, the Time to (collision risk) Zone (TTZ) can be calculated using

$$TTZ_v = \frac{d_v}{v_v} \qquad TTZ_p = \frac{d_p}{v_p}$$

where $TTZ_v$ and $TTZ_p$ are, respectively, vehicle and pedestrian time to zones. After the TTZs are calculated, a basic prediction algorithm was implemented to compare these values and determine if there is a collision using

$$TTZ_p - t_s < TTZ_v < TTZ_p + t_s$$

where $t_s$ is a time safety margin, a parameter that should be determined as a constant or dynamic value, considering the driver preferences as well as Global Positioning System (GPS) accuracy. In this study, it was chosen as a constant value of 4. With this logic, there is a possible collision when vehicle and pedestrian are approaching the collision zone at the same time. Although a simple algorithm was implemented to handle collision prediction within the scope of this paper, prediction can be improved using more sophisticated methods such as Kalman filtering [16] or neural networks [17].

## Driver Warning

The application running inside the vehicle phone handles the calculations for collision possibility and should determine how to warn the driver according to this calculation. In this study, we implemented a warning system with various severity degrees to convey the degree of danger to the driver/car, hence, helping the driver's decision on how to slow down and stop. The main criteria in determining the severity of the warning was selected to be the vehicle TTZ, since the majority of the speed adjustment is expected to be on the vehicle side. While the TTZ gets smaller, warning severity increases.

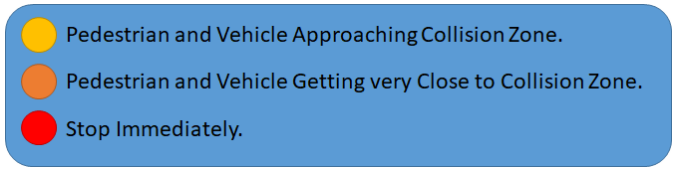

🟡 Pedestrian and Vehicle Approaching Collision Zone.

🟠 Pedestrian and Vehicle Getting very Close to Collision Zone.

🔴 Stop Immediately.

Figure 3. Warning severities and meanings.

Three severities of warnings were determined for the implementation, along with the fourth state where there is no warning issued. These three severities were shown with their descriptions next to them in Figure 3. They will be addressed with their color within the remainder of this paper. Vehicle starts with no warning even if there is a possible collision in distant future. When the collision becomes closer in terms of time, where $TTZ_v$ gets smaller than a certain threshold, the yellow warning is issued. The yellow warning is the lightest severity where there is a possibility for collision but it can be avoided easily by reducing speed slowly and very comfortably. When the collision zone becomes closer and $TTZ_v$ gets smaller, then a second threshold, orange warning is issued. This means the vehicle has to start slowing down as soon as possible with a relatively uncomfortable amount of deceleration to be able to stop safely. These two thresholds can be chosen as constant values, similar to the time safety margin discussed in the previous section.

Third and final severity is red, where the vehicle has to brake immediately with harsh deceleration to be able to stop safely. Unlike the threshold values defined for the previous two severities, this value is dynamic due to vehicle speed and distance being variable. Therefore, it needs to be calculated every step to issue the driver this warning if necessary. We can calculate the minimum amount of deceleration $a_{min}$ to stop at a given distance to zone with given vehicle speed as

$$a_{min} = \frac{v_v^2}{2d_v}$$

If the minimum deceleration exceeds the value we have for the braking deceleration, it means we must immediately press full brake to stop the vehicle. Therefore, checking the condition,

$$a_{min} > a_{brake}$$

would yield the red warning trigger, where $a_{brake}$ is the full brake deceleration or the maximum amount of deceleration we would like to have. Keep in mind that this is the ideal condition where the driver is able to press the brakes with no delay and the vehicle is able to stop with the determined braking deceleration value on a straight road. Therefore, an extra margin can be added on top of this to cover the driver reaction time. If desired, there are also more involved formulas for stopping distance computation that take friction and road slope into account along with the driver reaction time [18].

After the driver warning is calculated using TTZ information as well as distance and speed information for the red warning, the warning is issued to the driver on the phone screen as a large and colored warning symbol with pedestrian indication. The warning sign also blinks with a fade in and out animation to ensure that the driver notices it.

## Testing and Results

The algorithm was implemented in an Android app along with Bluetooth communication as discussed in the previous section. The test location was selected specifically to create an NLOS scenario at an intersection with no traffic lights in order to demonstrate the capability of providing a remedy for this type of dangerous situation. This scenario is illustrated on top of a satellite image of the test location in Figure 4.

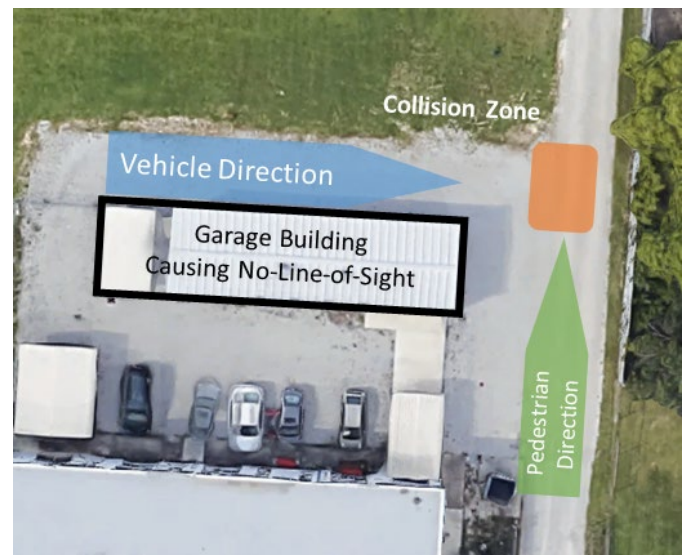

Figure 4. Test scenario illustration.

The vehicle starts around the starting edge of the indicated garage building and moves forward towards the intersection, indicated with a blue arrow, while the pedestrian starts behind the building and away from the intersection as indicated with the green arrow. The collision zone is also illustrated and was used to calculate TTZs. Two phones running the application were used in the test, one in the vehicle and the other one being carried by the pedestrian. Along with the data recording inside the phone for communication and warning calculation data, a camera was fixed inside the vehicle to record part of the driver view and the Head-Up Display (HUD) where the warning was displayed. Snapshots from the recorded video can be seen in Figure 5.

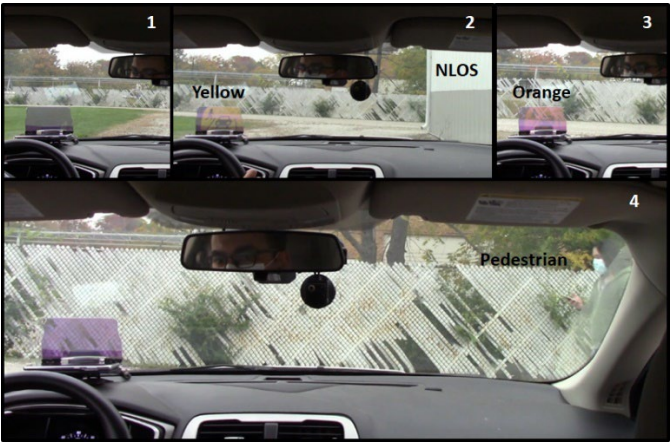

Figure 5. In vehicle snapshots from testing.

As seen in the figure above, a HUD was fixed in front of the driver where the phone is placed on with its screen facing upward. The reflection of the screen on the HUD glass can be observed clearly by the driver although it was a bit faded on the camera capture in Figure 5. Snapshots in Figure 5 are numbered according to the chronological order of events which correspond to the vehicle getting closer to the wall across the T intersection. The vehicle starts to drive with no warning in snapshot 1. Since the phone is receiving the P2V transmission even with no line of sight, as it can be seen in the snapshot 2, the driver receives a yellow warning, indicating that there is a pedestrian approaching towards the intersection. In snapshot 3, the vehicle comes closer to the intersection without slowing down and the warning turns to orange to indicate increased severity. Finally in snapshot 4, the vehicle stops safely as the pedestrian shows up on the right side of the windshield while starting to cross in front of the vehicle. The HUD window is clear with no warnings in this case since the vehicle has stopped. Hence, the testing is successfully completed for the application. The data recorded from the phone is shown in the color coded visualization of Figure 6.

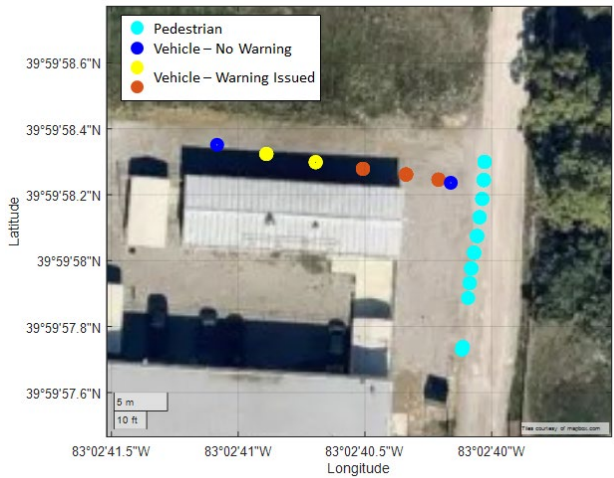

Figure 6. Visualization of communication and warning data.

The garage building can be seen as the gray rectangle in Figure 6 and the vehicle and pedestrian positions are plotted along as colored circles with the warning level indicated by the color used. Light blue represents the pedestrian location while the vehicle location is represented by dark blue color when there is no warning. When there is a warning, vehicle location is indicated by the color of the warning, which can be seen as yellow or orange circles on the graph. The whole experiment can be observed in detail using this visualization. The vehicle starts at the top left with no warning as the pedestrian starts at the bottom right. The location of points represents the GPS measurements where the vehicle was receiving the P2V messages. A while after the vehicle starts moving, the driver receives a yellow warning while the pedestrian is still behind the garage building showing that the method works successfully for the NLOS case. Following the yellow warning, an orange warning is issued as the vehicle gets closer to the collision zone without slowing down. Finally, the vehicle stops safely and the warning is disabled. This location can be seen as a blue circle. The pedestrian continues to move and crosses in front of the stopped vehicle.

## Summary/Conclusions

A pedestrian safety android application implementation using Bluetooth was discussed and demonstrated in this paper. The main goal of the application was to improve pedestrian safety by transmitting pedestrian information to vehicles, which allows a warning to the driver to be issued even in NLOS scenarios. The design consists of several modules, scenario configuration, collision prediction and driver warning calculation. These modules were implemented inside the Android application that uses GPS sensors to measure the necessary information and transmits it via Bluetooth. Contribution of the paper is the combination and method of implementation of these modules in a mobile application form, along with real-world testing of this application. The application was tested successfully with a vehicle and pedestrian at a testing area with NLOS. Test results successfully demonstrated the potential of the implementation and the use case. While the work here was presented as a driver warning system, this approach can also be integrated into an Advanced Driver Assistance System (ADAS) or an Autonomous Vehicle (AV) to automatically stop the vehicle or to execute an evasive maneuver if this cannot be done.

Use of this P2V communication based approach as part of such a pedestrian collision avoidance system will require the use of robust trajectory following controllers to handle uncertainty and disturbances [19], [20], [21], [22], [23], [24], [25], [26], [27] and can also benefit from yaw dynamics improvement [28], [29], evasive maneuvers [30] and integration with ADAS and other vehicle control systems [31]-[38].

In an actual future deployment of this approach, potential problems with multiple pedestrians broadcasting P2V messages and the effect of these warnings and automated actions on the rest of the road traffic should also be investigated.

It is important to emphasize that implementation in this paper is aimed for the demonstration of the chosen use case. Therefore, it is designed as a combination of basic modules where each can be modified or improved individually with more sophisticated algorithms to increase overall performance and safety. Moreover, measurement accuracy of mobile device sensors should be kept in mind. Device related limitations are expected to be solved with developments in sensor technology, such as the dual-frequency GPS sensors which provide up to a few decimeters of accuracy and are starting to have wider availability in recent mobile phone models. Although the application is limited to Android mobile phones in this study, IOS implementation can also be considered to cover a wider variety of VRUs. These improvements are planned to be part of the future work.

# References

[1] S. Y. Gelbal, M. R. Cantas, B. A. Guvenc, L. Guvenc, G. Surnilla, H. Zhang, M. Shulman, A. Katriniok and J. Parikh, "Hardware-in-the-Loop and Road Testing of RLVW and GLOSA Connected Vehicle Applications," in *SAE WCX*, 2020.

[2] I. Vourgidis, L. Maglaras, A. S. Alfakeeh, A. H. Al-Bayatti and M. A. Ferrag, "Use Of Smartphones for Ensuring Vulnerable Road User Safety through Path Prediction and Early Warning: An In-Depth Review of Capabilities, Limitations and Their Applications in Cooperative Intelligent Transport Systems," *Sensors,* vol. 20, no. 4, p. 997, 2020.

[3] S. Gelbal, B. Aksun-Güvenç and L. Guvenc, "Elastic Band Collision Avoidance of Low Speed Autonomous Shuttles with Pedestrians," *International Journal of Automotive Technology,* vol. 21, no. 4, pp. 903-917, 2020.

[4] S. Gelbal, S. Arslan, H. Wang, B. Aksun-Güvenç and L. Güvenç, "Elastic Band Based Pedestrian Collision Avoidance using V2X Communication," in *IEEE Intelligent Vehicles Symposium*, Redondo Beach, California, 2017.

[5] M. Bagheri, M. Siekkinen and J. K. Nurminen, "Cellular-based vehicle to pedestrian (V2P) adaptive communication for collision avoidance," in *International Conference on Connected Vehicles and Expo (ICCVE)*, 2014.

[6] P. Ho and J. Chen, "WiSafe: Wi-Fi Pedestrian Collision Avoidance System," *IEEE Transactions on Vehicular Technology,* vol. 66, no. 6, pp. 4564-4578, 2017.

[7] X. Wu, R. Miucic, S. Yang, S. Al-Stouhi, J. Misener, S. Bai and W.-h. Chan, "Cars Talk to Phones: A DSRC Based Vehicle-Pedestrian Safety System," in *IEEE 80th Vehicular Technology Conference*, 2014.

[8] J. Kotte, C. Schmeichel, A. Zlocki, H. Gathmann and L. Eckstein, "Concept of an enhanced V2X pedestrian collision avoidance system with a cost function–based pedestrian model," *Traffic Injury Prevention,* vol. 18, no. sup1, pp. S37-S43, 2017.

[9] S. Schwarz, D. Sellitsch, M. Tscheligi and C. Olaverri-Monreal, "Safety in pedestrian navigation: road crossing habits and route quality needs," in *3rd International Symposium on Future Active Safety Technology Toward zero traffic accidents*, 2015.

[10] Q. Wang, B. Guo, G. Peng, G. Zhou and Z. Yu, "CrowdWatch: pedestrian safety assistance with mobile crowd sensing.," in *ACM International Joint Conference*, 2016.

[11] H. Chu, V. Raman, J. Shen, A. Kansal, V. Bahl and R. R. Choudhury, "I am a smartphone and I know my user is driving," in *Sixth International Conference on Communication Systems and Networks (COMSNETS)*, 2014.

[12] H. Wang, A. Tota, B. Aksun-Guvenc and L. Guvenc, "Real time implementation of socially acceptable collision avoidance of a low speed autonomous shuttle using the elastic band method," *Mechatronics,* vol. 50, pp. 341-355, 2018.

[13] L. Guvenc, B. Aksun-Guvenc, S. Zhu and S. Gelbal, Autonomous Road Vehicle Path Planning and Tracking Control, New York: Wiley / IEEE Press, Book Series on Control Systems Theory and Application, 2002.

[14] SAE, "J2735 Dedicated Short Range Communications (DSRC) Message Set Dictionary," 2016.

[15] "Bluetooth Low Energy Advertising," 29 October 2021. [Online]. Available: https://source.android.com/devices/bluetooth/ble_advertising.

[16] C. Yu, L. Haiyong, Z. Fang, W. Qu and S. Wenhua, "Adaptive Kalman filtering-based pedestrian navigation algorithm for smartphones," *International Journal of Advanced Robotic Systems,* 2020.

[17] S. Zamboni, Z. T. Kefato, S. Girdzijauskas, C. Norén and L. D. Col, "Pedestrian trajectory prediction with convolutional neural networks," *Pattern Recognition,* vol. 121, 2022.

[18] American Association of State Highway and Transportation Officials (AASHTO), A Policy on Geometric Design of Highway and Streets, Washington, DC, 1994.

[19] H. Wang, S. Gelbal and L. Guvenc, ""Multi-Objective Digital PID Controller Design in Parameter Space and its Application to Automated Path Following," *IEEE Access,* vol. 9, pp. 46874-46885, 2021.

[20] B. Demirel and L. Guvenc, "Parameter Space Design of Repetitive Controllers for Satisfying a Mixed Sensitivity Performance Requirement," *IEEE Transactions on Automatic Control,* vol. 55, pp. 1893-1899, 2010.

[21] B. Aksun-Guvenc and L. Guvenc, "Robust Steer-by-wire Control based on the Model Regulator," in *IEEE Conference on Control Applications*, 2002.

[22] B. Orun, S. Necipoglu, C. Basdogan and L. Guvenc, "State Feedback Control for Adjusting the Dynamic Behavior of a Piezo-actuated Bimorph AFM Probe," *Review of Scientific Instruments,* vol. 80, no. 6, 2009.

[23] L. Guvenc and K. Srinivasan, "Friction Compensation and Evaluation for a Force Control Application," *Journal of Mechanical Systems and Signal Processing,* vol. 8, no. 6, pp. 623-638.

[24] M. Emekli and B. Aksun-Guvenc, "Explicit MIMO Model Predictive Boost Pressure Control of a Two-Stage Turbocharged Diesel Engine," IEEE Transactions on Control Systems Technology, vol. 25, no. 2, pp. 521-534, 2016.

[25] Aksun-Guvenc, B., Guvenc, L., 2001, "Robustness of Disturbance Observers in the Presence of Structured Real Parametric Uncertainty," Proceedings of the 2001 American Control Conference, June, Arlington, pp. 4222-4227.

[26] Guvenc, L., Ackermann, J., 2001, "Links Between the Parameter Space and Frequency Domain Methods of Robust Control," International Journal of Robust and Nonlinear Control, Special Issue on Robustness Analysis and Design for Systems with Real Parametric Uncertainties, Vol. 11, no. 15, pp. 1435-1453.

[27] Demirel, B., Guvenc, L., 2010, "Parameter Space Design of Repetitive Controllers for Satisfying a Mixed Sensitivity Performance Requirement," IEEE Transactions on Automatic Control, Vol. 55, No. 8, pp. 1893-1899.

[28] Guvenc, L.; Aksun-Guvenc, B. The Limited Integrator Model Regulator and Its Use in Vehicle Steering Control. Turkish Journal of Engineering and Environmental Sciences 2002, 26, 473–482.

[29] Aksun-Guvenc, B., Guvenc, L.; Ozturk, E.S.; Yigit, T. Model Regulator Based Individual Wheel Braking Control. In Proceedings of 2003 IEEE Conference on Control Applications, 2003. CCA 2003.; June 2003; Vol. 1, pp. 31–36.

[30] Ding, Y., Zhuang, W., Wang, L., Liu, J., Guvenc, L., Li, Z., 2020, "Safe and Optimal Lane Change Path Planning for Automated Driving," IMECHE Part D Passenger Vehicles, Vol. 235, No. 4, pp. 1070-1083, doi.org/10.1177/0954407020913735.

[31] T. Hacibekir, S. Karaman, E. Kural, E. S. Ozturk, M. Demirci and B. Aksun Guvenc, "Adaptive headlight system design using hardware-in-the-loop simulation,"2006 IEEE International Conference on Computer Aided Control System Design, 2006 IEEE International Conference on Control Applications, 2006 IEEE International Symposium on Intelligent Control, Munich, Germany, 2006, pp. 915-920, doi: 10.1109/CACSD-CCA-ISIC.2006.4776767.

[32] Guvenc, L., Aksun-Guvenc, B., Emirler, M.T. (2016) "Connected and Autonomous Vehicles," Chapter 35 in Internet of Things/Cyber-Physical Systems/Data Analytics Handbook, Editor: H. Geng, Wiley.

[33] Emirler, M.T., Uygan, I.M.C., Aksun-Guvenc, B., Guvenc, L. (2014) "Robust PID Steering Control in Parameter Space for Highly Automated Driving," International Journal of Vehicular Technology, Vol. 2014, Article ID 259465.

[35] Emirler, M.T.; Guvenc, L.; Guvenc, B.A. Design and Evaluation of Robust Cooperative Adaptive Cruise Control Systems in Parameter Space. International Journal of Automotive Technology 2018, 19, 359–367, doi:10.1007/s12239-018-0034-z.

[36] Gelbal, S.Y.; Aksun-Guvenc, B.; Guvenc, L. Collision Avoidance of Low Speed Autonomous Shuttles with Pedestrians. International Journal of Automotive Technology 2020, 21, 903–917, doi:10.1007/s12239-020-0087-7.

[37] Zhu, S.; Gelbal, S.Y.; Aksun-Guvenc, B.; Guvenc, L. Parameter-Space Based Robust Gain-Scheduling Design of Automated Vehicle Lateral Control. IEEE Transactions on Vehicular Technology 2019, 68, 9660–9671, doi:10.1109/TVT.2019.2937562.

[38] Yang, Y.; Ma, F.; Wang, J.; Zhu, S.; Gelbal, S.Y.; Kavas-Torris, O.; Aksun-Guvenc, B.; Guvenc, L. Cooperative Ecological Cruising Using Hierarchical Control Strategy with Optimal Sustainable Performance for Connected Automated Vehicles on Varying Road Conditions. Journal of Cleaner Production 2020, 275, 123056, doi:10.1016/j.jclepro.2020.123056.

## Contact Information

Will be entered after the review process.

## Acknowledgments

The Ohio State University authors would like to thank Ford Motor Company for partial support of this work.

## Definitions/Abbreviations

| | |
|---|---|
| **VRU** | Vulnerable Road User |
| **V2X** | Vehicle to Everything |
| **V2V** | Vehicle to Vehicle |
| **V2I** | Vehicle to Infrastructure |
| **V2P** | Vehicle to Pedestrian |
| **DSRC** | Dedicated Short Range Communication |
| **C-V2X** | Cellular Vehicle to Everything |
| **PSM** | Personal Safety Message |
| **P2V** | Pedestrian to Vehicle |
| **NLOS** | No-Line-of-Sight |
| **API** | Application Programming Interface |
| **LOS** | Line-of-Sight |
| **BLE** | Bluetooth Low Energy |
| **NHTSA** | National Highway Traffic Safety Administration |
| **TTZ** | Time to Zone |
| **GPS** | Global Positioning System |
| **HUD** | Head-Up Display |